Technical report preprint.

# GRIN Transfer: A production-ready tool for libraries to retrieve digital copies from Google Books.

**Liza Daly** [a], **Matteo Cargnelutti** [a], **Catherine Brobston** [a], **John Hess** [b],
**Greg Leppert** ✉ [a], **Amanda Watson** [c], **Jonathan Zittrain** [d]

[a] Institutional Data Initiative, Harvard Law School Library
[b] Library Innovation Lab, Harvard Law School Library
[c] Harvard Law School Library
[d] Harvard Law School, Harvard School of Engineering and Applied Sciences, Harvard Kennedy School

## Abstract

Publicly launched in 2004, the Google Books project has scanned tens of millions of items in partnership with libraries around the world. As part of this project, Google created the Google Return Interface (GRIN). Through this platform, libraries can access their scanned collections, the associated metadata, and the ongoing OCR and metadata improvements that become available as Google reprocesses these collections using new technologies. When downloading the Harvard Library Google Books collection from GRIN to develop the Institutional Books dataset, we encountered several challenges related to rate-limiting and atomized metadata within the GRIN platform. To overcome these challenges and help other libraries make more robust use of their Google Books collections, this technical report introduces the initial release of **GRIN Transfer**. This open-source and production-ready Python pipeline allows partner libraries to efficiently retrieve their Google Books collections from GRIN. This report also introduces an updated version of our **Institutional Books 1.0 pipeline**, initially used to analyze, augment, and assemble the **Institutional Books 1.0 dataset**. We have revised this pipeline for compatibility with the output format of GRIN Transfer. A library could pair these two tools to create an end-to-end processing pipeline for their Google Books collection to retrieve, structure, and enhance data available from GRIN. This report gives an overview of how GRIN Transfer was designed to optimize for reliability and usability in different environments, as well as guidance on configuration for various use cases.

✉ Corresponding author: gleppert@law.harvard.edu

# 1 Contributions

With this technical report, we introduce the following contributions:

1. **The GRIN Transfer Pipeline**.
   An open-source and production-ready Python pipeline for libraries to efficiently retrieve their Google Books collection from GRIN.
   Beyond automating the retrieval and storage of volumes from a given Google Books collection in a workflow-agnostic way, this pipeline allows for extracting and organizing the bibliographic and processing-related metadata for each volume, as well their OCR-extracted text.
   This pipeline is compatible with GRIN at the time of its release (November 2025), and is available at: https://github.com/institutional/grin-transfer.
2. **Updated Institutional Books 1.0 Pipeline.**
   Alongside GRIN Transfer, we introduce an updated version of the pipeline we used to analyze, augment and release the Institutional Books 1.0 dataset (Cargnelutti et al., 2025), derived from Harvard Library's Google Books collection.
   This pipeline, now fully compatible with the output of GRIN Transfer, can be used by libraries in tandem with GRIN Transfer to process their Google Books collections as data. The pipeline will analyze and augment the collection's metadata (e.g: language detection, topic classification), post-process the OCR-extracted text to improve usability, and identify likely duplicates at the collection-level.
   Available at: https://github.com/institutional/institutional-books-1-pipeline
3. **Insights into Google Books and GRIN.**
   In the process of developing GRIN Transfer, we gathered insights into the challenges and opportunities of working with Google Books' return interface at collection-scale, as observed at the time of developing this initial version of the pipeline.

# 2 Background

Publicly launched in 2004, the Google Books project has scanned tens of millions of items from library collections around the world (Lee, 2019). As one of the earliest Google Books scanning partners, Harvard Library is among its largest contributors. This partnership paved the way for the Institutional Data Initiative (IDI) at Harvard Law School Library to release the Institutional Books dataset in June 2025. Institutional Books comprises the entirety of Harvard Library's public domain Google Books collection – roughly 1 million volumes scanned to date – and was created with the goal of making large scale, high quality data available for computational research and AI training.

The first step in the multiyear process of releasing this dataset was to download Harvard Library's digital copies of the collection from the Google Return Interface (GRIN), Google's access platform for libraries. Through this platform, libraries can download their scanned collections, the associated metadata, and the ongoing OCR and metadata improvements that become available as Google reprocesses these collections using new technologies.

This initial download process spanned approximately three months and required an engineering resource to work through various challenges. These included discovering, investigating, and accounting for platform-wide rate limiting constraints (Section 3, Section 4, Section 5), as well as piecing together disparate sources of metadata (Section 4.6). The GRIN platform is maintained by Google with the goal of facilitating each library's access to their scanned materials, but most partner libraries have yet to take full advantage of this access. The reasons for this are diverse, but challenges like the ones we experienced are a contributing factor.

After completing the download process, the IDI team committed to releasing **GRIN Transfer**—a toolchain to enable other libraries to efficiently replicate our work for their own collections. We spoke to several libraries who have, through the decades, written their own tools to interact with GRIN. While these tools support some of the functionality of GRIN Transfer, the libraries often described them as difficult to maintain or limited in scope. They welcomed a modern and actively maintained tool to support their ongoing use of these collections. Our hope is that GRIN Transfer can meet this need and, in so doing, help libraries make more robust use of their own digitized collections and reap the full benefits of their participation in Google Books.

Alongside GRIN Transfer, we are releasing an updated version of our **Institutional Books 1.0 Pipeline** to seamlessly integrate with the output from GRIN Transfer. A library could pair GRIN Transfer with the newly compatible Institutional Books 1.0 Pipeline to create an end-to-end processing workflow for their Google Books collections that would retrieve, structure, and enhance the scanned images and data available from GRIN, making them even more usable for library purposes.

## 3 Design

GRIN Transfer was designed to have a small operational footprint and run as a command-line tool in most modern Linux-like environments. With that frame in mind, we chose to focus on relative genericity, portability, and ease of use:
- The pipeline supports Python 3.12 and above. For users with older Python installations, we supplied a ready-to-use Docker (Boettiger, 2015) configuration.
- To further optimize for portability, GRIN Transfer tracks its work using a SQLite database, which was chosen for its limited footprint and high interoperability (Gaffney et al., 2022; sqlite.org, n.d.). Because SQLite databases are file-based, they can easily be copied, transferred and archived, which GRIN Transfer periodically does. Internally, and as part of our development process, we used this database to write custom reporting tools; example queries for common use cases are provided in GRIN Transfer's documentation[1].
- To provide visibility into the total book catalog, GRIN Transfer can output a CSV of all volume metadata and a machine-readable summary of pipeline executions to date. The CSV format was chosen for its simplicity and interoperability (Wilson et al., 2017).
- Because even modest institutional Google Books collections can require significant storage capacity to download, GRIN Transfer is intended to be agnostic about storage architecture. Based on user feedback, we chose to support a locally-addressable filesystem and any S3-compatible block storage layer as final storage options.

Working with GRIN entails managing and monitoring resource constraints, which we aimed to streamline. Per Google's documentation and direct observation, the most salient constraints we observed were a global five (5) queries-per-second rate limit on all GRIN requests, and a maximum length of 50,000 queued conversion requests. At the time of writing this technical report, synchronizing a single volume from a cold start requires—at minimum—four distinct GRIN requests:

---
[1] https://github.com/institutional/grin-transfer/blob/main/docs/sql-queries.md

1. an initial collection of the volume to retrieve its unique barcode and basic metadata
2. a request to convert the volume for download
3. a request to download the volume (once converted)
4. a detailed metadata request to complete the record.

Because GRIN does not currently offer a notification or PubSub framework (Eugster et al., 2003), download readiness can only be assessed by polling GRIN periodically. In practice, a single file package download requires many more than those four requests over its lifecycle. GRIN Transfer was designed to account for this constraint.

We also designed GRIN Transfer to be unopinionated about the workflow that a user or institution would implement. Some institutions have never worked with GRIN before; others have fully downloaded their entire collection in the past and need to do so again. Some institutions may want a one-time snapshot; others ongoing collection management. To that end, GRIN Transfer organizes work as composable "queues" of barcodes that represent volumes in some initial state. It also supports arbitrary volume identifier (barcode) lists which could be fed from external systems.

A full synchronization of a collection can run over the course of days or weeks (Section 5). Therefore, it was imperative that GRIN Transfer be resilient during long-running operations. It must be able to resume cleanly after interruption, output adequate logs and metrics to assess progress, and recover from transient network or storage errors.

## 4 Mode of operation

GRIN Transfer operates through a series of workflows, some invoked multiple times to fully synchronize an entire collection (Figure 1).

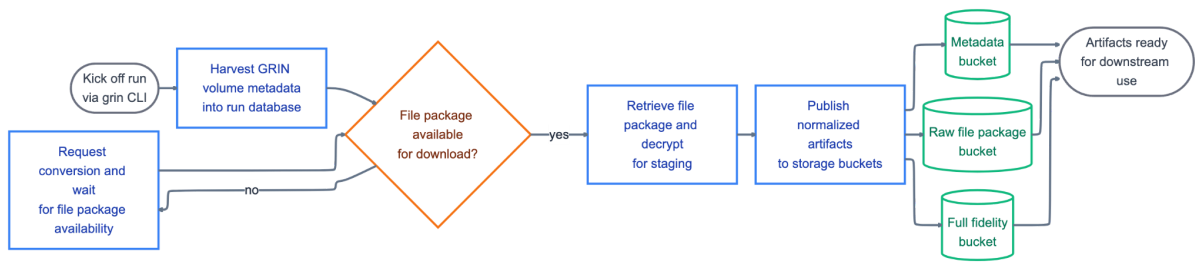

*Figure 1: Flowchart diagram.*
Overview of GRIN Transfer's mode of operation.

It begins its processing pipeline with the collection step, during which all volume metadata is downloaded from GRIN and stored locally for evaluation and further processing. During our tests, collecting all book metadata for Harvard Library's collection of over one million volumes could be completed within an hour.

Before volume scans could be downloaded, GRIN requires that they be converted into "file packages"—encrypted, compressed archives that include metadata, full-text OCR, and page scans. Collectively, GRIN Transfer refers to this set of data as the "raw" file for a volume. If a file package is not currently available for a given volume, GRIN provides a URL to which the volume's uniquely identifying barcode can be submitted. GRIN will then attempt to convert the volume for download asynchronously.

Once available for download, GRIN Transfer can retrieve the encrypted file package, decrypt it using the institution's decryption key, perform any requested intermediate processing, and upload or copy the file to the desired long-term storage location.

## 4.1 Configuration

Software operators of GRIN Transfer need to supply three parameters to connect to and download from GRIN:

- **Their GRIN Google Account.**
  This is a Google user account that was supplied to a Google partner manager in order to access GRIN. The operator authorizes this account to use GRIN Transfer via OAuth 2.0.
- **Their institution's Library Directory.**
  This is a unique identifier, supplied by Google and used by GRIN to segment the institution's collection. The Library Directory appears in authenticated URLs as `https://books.google.com/libraries/LIBRARY_DIRECTORY/`, e.g. `https://books.google.com/libraries/Harvard/`. This value is case-sensitive.
- **GRIN archive decryption key.**
  Volume archives stored in GRIN are encrypted. A GPG decryption passphrase, supplied by Google, is required to successfully unpack and read book archives.

GRIN Transfer obtains these credentials from the user by retrieving them from the operator's home directory in `~/.config/grin-transfer`.

## 4.2 Pipeline state flow

GRIN Transfer organizes its workloads into "queues": lists of barcodes that are collectively in some known state. One or more queues can be processed in a single invocation and in a desired order.

The pre-configured queues are:
1. **Unconverted**: these are volumes that have never been converted into file packages for download. This is the default state for a barcode that has never been downloaded, either from this tool or from the GRIN web UI. Unconverted volumes can not be downloaded—they need to be processed by GRIN first into a downloadable file package. In GRIN, these volumes have a blank status field.
2. **Converted**: these volumes have been processed by GRIN and are ready to be downloaded. In GRIN, these volumes have the status `CONVERTED`. Once converted, GRIN keeps file packages available for download for approximately two weeks, per Google's documentation.
3. **Previous**: these volumes have been downloaded previously outside of GRIN Transfer (either manually from the GRIN web UI, or through some other automation). In GRIN, these volumes have the status of `PREVIOUSLY_DOWNLOADED`. The `previous` queue is intended for use by institutions who need to make a comprehensive snapshot regardless of prior activity.

Because GRIN Transfer cannot predict all possible ways to segment volumes, operators can supply a list of arbitrary barcodes either directly on the command line or by supplying a text file containing one barcode per row. This list of barcodes can be generated from an external system such as library management software, or through custom SQL queries against GRIN Transfer's SQLite database.

### 4.3 Collection inventory

GRIN Transfer manages the internal state of each volume in its local SQLite database, mirroring the state gathered from GRIN's "All Books" UI. In the initialization phase of GRIN Transfer, the operator runs `grin.py collect` to automatically page through all GRIN results and locally mirror its list of barcodes and metadata. The operation could be re-run at any time to collect new volumes or update existing metadata. A detailed breakdown of this process is available in Appendix A.

### 4.4 Conversion request

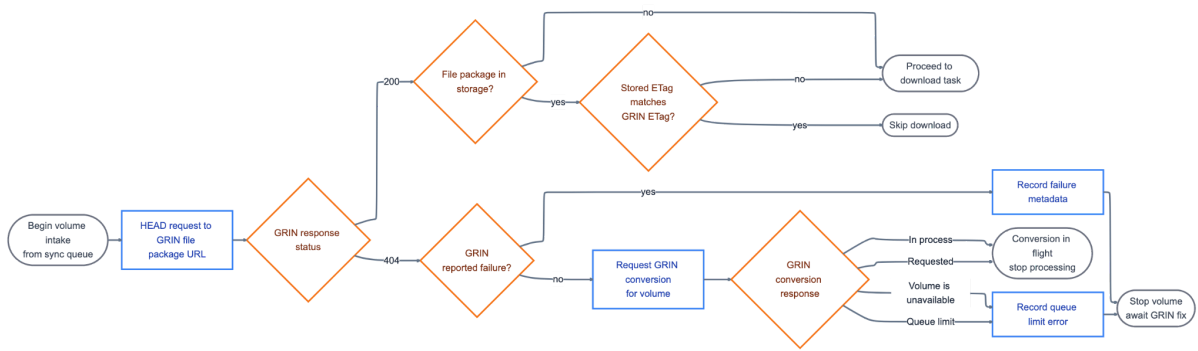

*Figure 2: Flowchart diagram.*
GRIN Transfer's conversion request process.

Volume file packages cannot be directly downloaded from GRIN until they are converted by Google Books' backend (Figure 2). GRIN allows users to send one or more barcodes to a URI to request this conversion. Based on our observation during the re-conversion of Harvard Library's collection, approximately 80% of such requests are fulfilled within 48 hours.

GRIN does not currently provide a mechanism to notify when a volume has been converted. After the initial conversion request is sent, GRIN needs to be queried again to check if the `state` field of the volume changed to `CONVERTED`, indicating that the file package is available. If an error was encountered during conversion, GRIN lists the error on its `failures` page. GRIN Transfer can use that information in subsequent runs to annotate the volume record with error details and operators can decide whether to take action. This could include re-queuing at a later time or attempting to correct the defect.

GRIN currently imposes a hard limit of 50,000 barcodes that can be present in the conversion queue. Subsequent requests to the "request conversion" endpoint are rejected with HTTP 429 "Too Many Requests." GRIN Transfer was designed to detect this condition and cease feeding new volumes to that queue during that particular run (Figure 2).

### 4.5 File package retrieval

Volume file packages in GRIN, if available for download, are present at a predictable URL:

`https://books.google.com/libraries/<LIBRARY_DIRECTORY>/<BARCODE>.tar.gz.gpg`.

GRIN Transfer leverages this predictability to proactively probe these URLs when seeking to download a volume. Based on the HTTP response codes and metadata returned from these requests, the workflow can branch into different tasks to fulfill the operator's request and sync the volume efficiently (Figure 2).

### 4.5.1 File package versioning

GRIN Transfer attempts to sync a volume by first issuing an HTTP HEAD request to the volume's file package download URL. An HTTP 200 response to that HEAD request indicates that the file package is available for download. GRIN provides additional metadata in the header of the response, including the file's ETag value (Fielding, Nottingham and Reschke, 2022), which uniquely identifies the version of the package.

GRIN Transfer stores this value in its local database. If the volume has previously been synced by GRIN Transfer, the ETag from the request is compared with the stored ETag. If the values match, it is inferred that the file package has not been changed since it was last synced and the retrieval task is skipped.

If the ETag does not match, or if the volume has never been synced, GRIN Transfer issues a GET request to the same URL. This triggers a download of the file package.

### 4.5.2 Decryption and unpacking

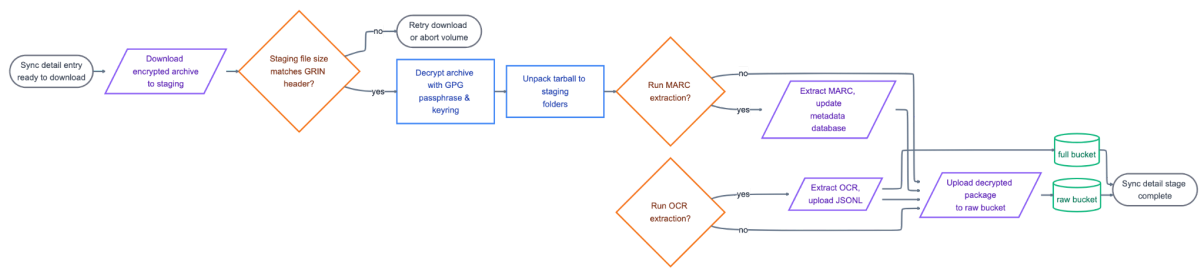

*Figure 3: Flowchart diagram.*
GRIN Transfer's volume decryption and unpacking process.

GRIN Transfer maintains a local staging directory where downloaded files are kept temporarily before being moved into the final storage location. Because we assume that the staging area has limited capacity relative to the size of the total corpus, GRIN Transfer continuously monitors available disk space and warns if a configurable threshold of 90% is exceeded. File packages are cleaned up after storage upload is completed; in most cases when the disk space threshold is exceeded, the pipeline can pause new downloads and the problem self-corrects.

Once downloaded, GRIN Transfer uses the institution's GPG key to decrypt the volume file package. After this, the file package is ready for additional processing or to be uploaded to archival storage (Figure 3).

If the optional content extraction tasks were requested, the archive will be automatically unpacked in the staging directory at the end of this phase.

### 4.4.3 MARC XML extraction

As an optional step, the pipeline can read the volume's packaged METS XML file (The Library of Congress, 2025), which contains MARC21 metadata (The Library of Congress, 2007) about the volume (Figure 3). The volume's metadata is stored in GRIN Transfer's tracking database and is materialized in the CSV export that is created with each run.

### 4.5.4 OCR text extraction

As another optional step, the pipeline can collate all of the discrete per-page OCR text files from a given volume and output them as a JSONL file (jsonlines.org, n.d.), with one page per line (Figure 4). This artifact was designed to be straightforward and efficient to consume by downstream full-text pipelines.

```
"\n"
"PE\n128\n55\nHARVARD COLLEGE\nVERI\nTAS\nSCIENCE CENTER\nLIBRARY\n"
""
"GE\n12\nE\n"
"1\n"
"E\n2\n"
"\nDEPARTMENT OF THE INTERIOR\nFRANKLIN K. LANE, Secretary\nUNITED STATES
GEOLOGICAL SURVEY\nGEORGE OTIS SMITH, Director\nBulletin 597\nGEOLOGY OF
  MASSACHUSETTS\nAND RHODE ISLAND\nBY\nB. K. EMERSON\nMENT OF THE
INT\nDEPART\nNTERIOR\nWASHINGTON\nGOVERNMENT PRINTING OFFICE\n1917\n"
"\nDELVBLWELL OF THE INLEKЮB\n65\nЕВУИКГТИ К ГКИЕ 26GLOFSLA\nJUTATE
ONTINU\nGEORGE CL\nTed aiseliull\nVAD KHODE 12Гуир\nθΕΟΓΟΟΣ ΟΕ
VERYCHELLE\nSCIENCE CENTER LIB.
\nTH\nBEEWEEBOЙ\n12\nПОЛЕТАМЕНТ ВЙСЬAG OLLION\n"
```

*Figure 4: JSONL excerpt*
GRIN Transfer's export of a given volume's OCR data as JSONL.

### 4.5.5 Upload

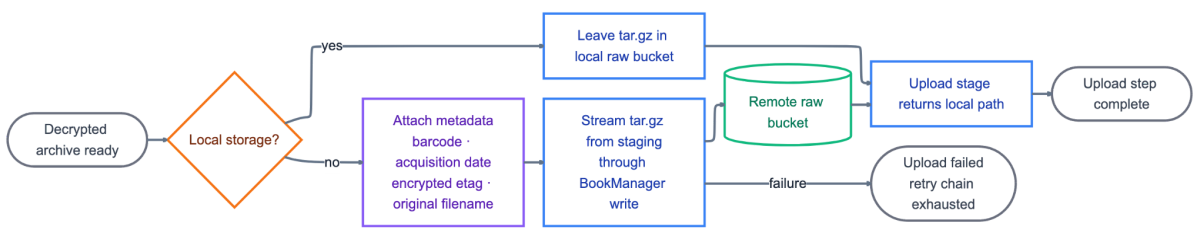

*Figure 5: Flowchart diagram.*
GRIN Transfer's data upload process.

The last step in the sync pipeline uploads the *decrypted* file package to storage (Figure 5). If the operator selected a local file system as the storage solution, the file is copied from the staging area to the storage directory. If OCR text extraction was performed, its JSONL artifact is uploaded to storage along with the decrypted file copy. To maximize throughput and reliability with large file packages, uploads to block storage are broken up into parts and uploaded concurrently.

Because S3-compatible block storage options generally support the association of object metadata with each entry, file packages are archived and annotated with the ETag of the *encrypted* file. This is used to optimize the download flow and allow skipping file packages that are byte-for-byte identical with those already in storage. Attaching the encrypted ETag to the object metadata means that the skip-if-identical process is resilient to the loss or corruption of the SQLite tracking database—potentially saving days or weeks of wall clock time re-syncing identical archives unnecessarily.

## 4.6 Metadata enrichment

GRIN does not present all volume metadata in a single endpoint; a subset of metadata about scanning and OCR quality is only available from the per-barcode search results page or from an unpaginated text-only endpoint unsuitable for large collections (Table 1). As an optional step, GRIN Transfer can request this metadata for all barcodes in the catalog; this process is called metadata enrichment. Because GRIN allows multiple barcodes to be requested at once, enriching the entire corpus can be achieved reasonably quickly despite the application-wide rate limits we observed.

*Table 1: GRIN Transfer's metadata sources*

| Metadata Group | Example Fields | Source | Populated during |
| --- | --- | --- | --- |
| Core identifiers | Barcode, title, google books URL | GRIN books listing | Collection phase |
| Google processing fields and timestamps | Scanned date, converted date, downloaded date, processed date, analyzed date, ocr date, state | GRIN books listing | Collection phase |
| GRIN condition fields | Viewability, opted out, conditions, scannable, tagging, audit, material error %, overall error %, claimed, ocr analysis score, ocr gtd score, digitization method, check-in date, source library bibkey, allow download updated date, viewability updated date | GRIN book detail view | Enrichment phase |
| MARC Bibliographic Metadata | Control Number, Date Type/Date 1/Date 2, Language, LCCN, LC Call Number, ISBNs, OCLC Numbers, Title/Subtitles, Personal/Corporate/Meeting Authors, Subjects, Genres, General Notes | GRIN METS XML (MARC records inside downloaded archives) | Parsed during file package sync |

## 4.7 "Hands free" mode

We anticipate that most GRIN Transfer use will happen with unattended, scheduled runs. The software was therefore designed to be triggered by a cron job or any Linux-compatible job runner. It can also be directly integrated into a different Python job runner as a library (Appendix B).

When executed from the command line or via cron, GRIN Transfer uses a session lock mechanism to prevent potentially conflicting write updates to the SQLite database. This behavior means it is safe to invoke the application on a timed schedule (e.g. hourly)—if a previous run is still going, the subsequent run is skipped.

Operators can decide on the most appropriate way to feed candidate barcodes to GRIN Transfer based on their workflow and the state of their corpus, but some common scenarios were anticipated.

*First-time archiving*

Institutions whose collections are available via Google Books may never have downloaded file packages at all. For these users, it is recommended that they invoke the pipeline with two built-in queues, in this order: `unconverted` (which issues conversion requests for new volumes, up to the 50,000 limit) and `converted` (which downloads any volumes that have been converted from previous runs). Scheduling runs at least twice a day would ensure that there are always volumes in the conversion request queue. Letting that queue empty down to zero means there is unallocated capacity on Google's side; it is likely desirable to keep that list topped up.

*Archiving a previously-active catalog*

Some institutions have worked with GRIN using earlier tooling, or made active use of the GRIN web UI to download file packages. These institutions may benefit from using the `previous` queue, which selects volumes that GRIN marked as previously downloaded. Volumes in this state can sometimes still be available for immediate download, so this queue first checks whether a file package is available before requesting that it be converted. This queue can be combined with `unconverted`—the operational behavior is identical, only the selection process differs—and is best followed by `converted`, to pick up volumes that have been made available.

*On-going collection management*

Institutions with active scanning programs may benefit from periodic re-runs of the `collect` step, which finds new barcodes and updates existing barcode metadata.

In this scenario, a daily or weekly `collect` invocation combined with regular runs of `sync pipeline --queue unconverted --queue converted` will continually sync newly-added volumes.

If it is desirable to re-sync volumes based on changes in metadata, operators can:
- Set up periodic `collect` runs to gather current metadata.
- Using a `sqlite3` client, write custom SQL to define the criteria desired to re-sync volumes—for example, finding the `MAX` date from GRIN date metadata and comparing it to the last sync time for the volume.
- Output the result of that SQL query as a file of barcodes to be ingested by `sync pipeline --barcodes-file <FILE>`

# 5 Known issues and limitations

We observed that GRIN currently enforces a 50,000-item conversion queue; once the backlog hits that threshold, GRIN rejects further requests with HTTP 429 responses. The service also caps activity at five requests per second, which keeps throughput at about four concurrent downloads while one to three workers handle HEAD requests. Even small overruns of the rate limit trigger throttling; it is currently unknown to us if heavier or continued rate limit violations would result in service disruption or rejections.

GPG decryption creates a post-download bottleneck when hardware acceleration is unavailable, especially on hosts with limited CPU or I/O capacity. This may limit the ability for operators to use extremely low-resource environments to run pipelines; a modern CPU is strongly recommended.

File-based storage can help sync file archives more quickly without network overhead, but provides less resiliency. Block storage offers stronger versioning and backup features. Block storage also makes GRIN Transfer rebuilds easier because each object carries its ETag metadata, allowing existing files to be skipped. With file-based storage, those ETags live only in the database, making a functional SQLite database a requirement for resuming interrupted syncs without needless work.

Our testing only covered Harvard Library's collection, which might not match other deployments in age or scope. Harvard's collection had already been fully synchronized with an earlier version of GRIN Transfer and so the file packages we were working with were not in a fully naive, unsynced state. Only local filesystem storage and Cloudflare R2 were fully tested in a production capacity as other storage options were not available at sufficient scale.

While we are releasing GRIN Transfer alongside our updated processing pipeline for use by the community, we know that the storage, compute, and technical support required to run them can be prohibitive for many institutions. We are developing avenues to provide further support for such institutions, and encourage teams facing these hurdles to get in touch.

# 6 Discussion and future directions

In advance of releasing GRIN Transfer, we connected with several Google Books libraries who shared early feedback that allowed us to improve the tool and its documentation before release. We encourage other users of the tool to share questions and feedback as they begin working with the tool, so that we may continue to improve its usability.

While we believe GRIN Transfer greatly reduces the complexity and time burden of working with GRIN at scale, we see more opportunity for improvement. In particular, we hope to work with Google to support potential adjustments to improve the use of GRIN. In order of anticipated impact, these could include:

1. Preprocessing all volumes in advance and making them immediately available in the converted, file package form. Alternatively, lifting the 50,000 cap on conversion requests would make retrieving collections for the first time more approachable.
2. Staging the encrypted archives in an authenticated cloud filesystem rather than providing them as HTTP-mediated downloads.
3. To facilitate monitoring for new or refreshed files, an export of metadata otherwise available only in GRIN via a similar mechanism to #2.

4. Clarity on the metadata fields users should look at to determine if a book has been reprocessed in some way (e.g. if the OCR has been improved).

We understand that all teams and tools face trade-offs, and know that we do not have full insight into the potential overhead of the Google Books team making the above changes. By offering these recommendations, our goal is to spur discussion and consideration within Google and its partner libraries.

We are also developing several other processing tools for use in concert with GRIN Transfer. These include a second version of our Institutional Books processing pipeline to improve handling of nonstandard formats and non-English languages, and a separate pipeline to extract and describe images from scanned pages.

## Acknowledgements

The authors of this technical report would like to thank:
- Gordon Nicol, Tony Stuart, and Rob Cawston from the National Library of Scotland for sharing context for potential use cases and requirements for GRIN Transfer, and for standing up the tool to provide feedback.
- Mia Ridge and Rossitza Atanassova at the British Library for sharing context into the library's technical environment and potential use cases for GRIN Transfer.
- Esme Cowles and Cliff Wulfman at Princeton University Library for sharing their experience working with GRIN, and for reviewing our code and providing feedback.
- Marcus Bitzl at the Bavarian State Library for sharing context into the library's technical environment, their previous work extracting data from GRIN, and potential use cases for GRIN Transfer.
- Harvard Library for the opportunity to work with their collection.
- The Google Books team for its continual improvement of library data and supporting tools.

## Disclaimers

GRIN Transfer is designed to efficiently synchronize a collection of file packages against current versions available in GRIN. Once a collection is deemed to be a complete preservation archive, it should be moved to permanent, read-only storage. While in operation, GRIN Transfer will overwrite file packages even if provided by GRIN in a degraded form, or write incomplete file packages due to network instability or interrupted operation.

# Appendices

## Appendix A: Collection inventory process

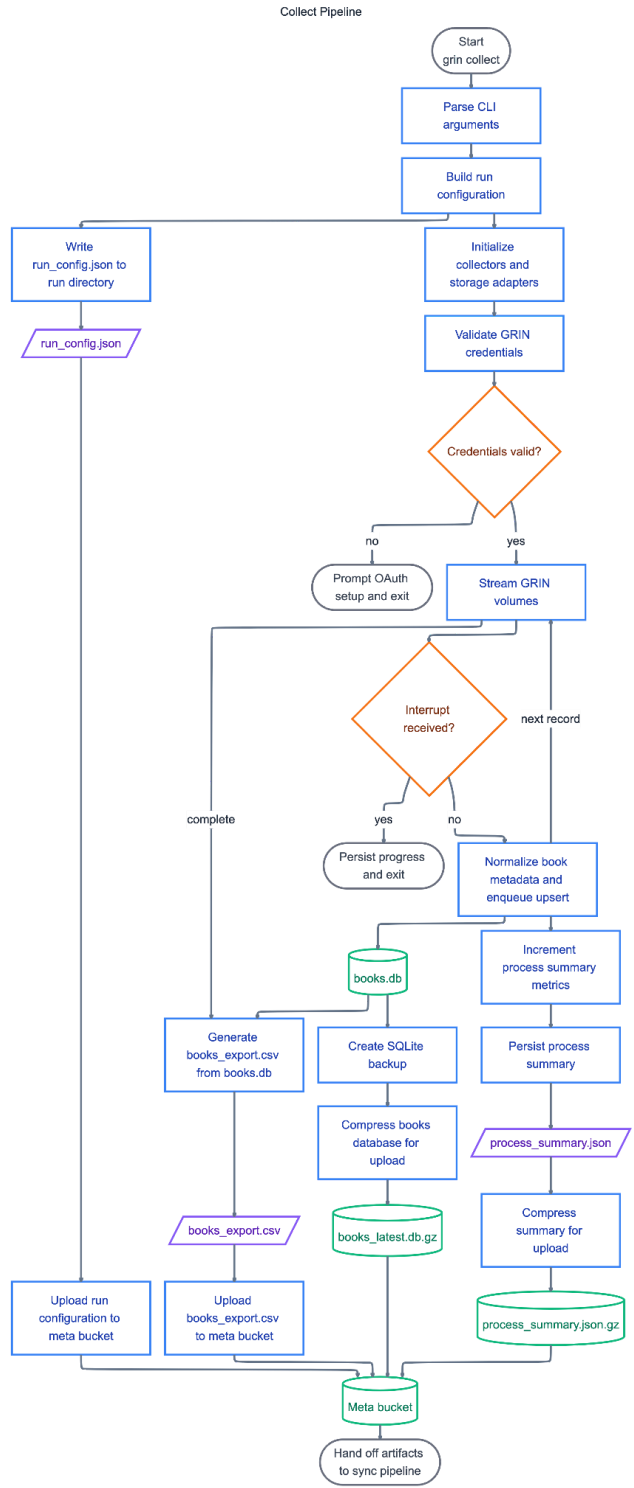

*Figure App.A: Diagram*
Flowchart diagram of GRIN Transfer's collection inventory process.

**Appendix B: Sample code for interacting with GRIN Transfer programmatically**

```python
async def sync_one_volume(barcode: str, config_file: Path):
    config = load_run_config(config_file)

    stage = ProcessStageMetrics("sync")
    stage.start_stage()

    pipeline = SyncPipeline.from_run_config(
        config=config,
        process_summary_stage=stage
    )

    await pipeline.initialize_resources()

    await pipeline.setup_sync_loop(
        queues=[],
        specific_barcodes=[barcode])

    await pipeline.cleanup()

    stage.end_stage()
```

*Figure App.B: Python code excerpt*
Example of how GRIN Transfer can be used programmatically in Python.